\documentclass[12pt]{article}
\pdfoutput=1

\usepackage{amsmath,amssymb,amscd, color, caption, cite, hyperref}

\usepackage{listings}
\usepackage{caption}
\usepackage{dsfont}
\usepackage{slashed}
\usepackage{color}
\usepackage{ulem}
\usepackage[pdftex]{graphicx}
\usepackage{epstopdf}
\usepackage{subfigure}
\usepackage{epsfig}
\usepackage{listings}
\usepackage{caption}
\usepackage{cite}
\usepackage{multirow}
\usepackage{comment}
\usepackage{hyperref}

\begin{document}

\baselineskip=21pt

\begin{center}

{\large {\bf Black Holes and Baryon Number Violation: Unveiling the Origins of Early Galaxies and the Low-Mass Gap}}

\vskip 0.2in
\vskip 0.2in

{\bf Merab Gogberashvili}\textsuperscript{a,b} and~
{\bf Alexander S.~Sakharov}\textsuperscript{c,d}~

\vskip 0.2in
\vskip 0.2in

{\small {\it

\textsuperscript{a}{\mbox Javakhishvili Tbilisi State University, 3 Chavchavadze Ave., Tbilisi 0179, Georgia}\\
\vspace{0.25cm}
\textsuperscript{b}{\mbox Andronikashvili Institute of Physics, 6 Tamarashvili St., Tbilisi 0177, Georgia}\\
\vspace{0.25cm}
\textsuperscript{c}{\mbox Department of Mathematics and Physics, Manhattan University,}\\
\mbox{4513 Manhattan College Parkway, Riverdale, NY 10471, United States of America}\\
\vspace{0.25cm}
\textsuperscript{d}{\mbox Experimental Physics Department, CERN, CH-1211 Gen\`eve 23, Switzerland}\\
\vspace{0.25cm}
}
}

\vskip 0.2in

{\bf Abstract}

\end{center}

\baselineskip=18pt
\noindent


We propose that modifications to the Higgs potential within a narrow atmospheric layer near the event horizon of an astrophysical black hole could significantly enhance the rate of sphaleron transitions, as well as transform the Chern-Simons number into a dynamic variable. As a result, sphaleron transitions in this region occur without suppression, in contrast to low-temperature conditions, and each transition may generate a substantially greater baryon number than would be produced by winding around the Higgs potential in Minkowski spacetime. This effect amplifies baryon number violation near the black hole horizon, potentially leading to a considerable generation of matter. Given the possibility of a departure from equilibrium during the absorption of matter and the formation of relativistic jets in supermassive black holes, we conjecture that this process could contribute to the creation of a significant amount of matter around such black holes. This phenomenon may offer an alternative explanation for the rapid growth of supermassive black holes and their surrounding galaxies in the early Universe, as suggested by recent observations from the JWST. Furthermore, this mechanism may provide insights into the low-mass gap puzzle, addressing the observed scarcity of black holes with masses near the Oppenheimer-Volkoff limit.

Early galaxies
\vskip 3mm

\noindent\today

\newpage


\section{Introduction} \label{intro}

One of the main challenges in cosmology and the Standard Model (SM) of particle physics is explaining the baryon asymmetry of the Universe \cite{Riotto:1999yt}. It is natural to consider that gravity may play an essential role in baryogenesis, particularly through its influence on fundamental interactions in a curved spacetime background \cite{Banks:2010zn}. Gravitationally induced particle creation in strong gravitational fields (an inherently out-of-equilibrium process) suggests that all of Sakharov’s conditions for baryogenesis might be satisfied near a black hole (BH) horizon \cite{Antunes:2019phe}. Indeed, previous studies have shown that BHs can trigger electroweak (EW) vacuum instability, both at zero temperature \cite{Burda:2015isa, Burda:2015yfa} and in the early universe \cite{Tetradis:2016vqb, Gorbunov:2017fhq, Canko:2017ebb, Mukaida:2017bgd, Kohri:2017ybt, Dai:2019eei}.

We propose that the Higgs potential may experience substantial modifications near a BH horizon, leading to an enhancement of baryon number-violating processes via sphaleron transitions. In this scenario, the BH is surrounded by a thin atmospheric layer of width $\epsilon$, generated by a $\delta$-source near the Schwarzschild horizon. This atmospheric concept builds on the "brick wall" \cite{tHooft:1984kcu}, which aimed to explain quantum effects near a BH horizon and provided a foundational framework for understanding boundary conditions in strong gravitational fields. Subsequent studies have extended this model, exploring its implications, particularly in the context of quantum field interactions near the BH horizon \cite{Ghosh:2009xg, brick1, brick2}.

In our study, we propose that this atmospheric layer forms within the intense gravitational field at a radial distance $r = 2M_{\rm BH} + \epsilon$, where the sphaleron decay rate is significantly enhanced, resembling conditions with unbroken EW symmetry. This scenario leads to a high frequency of baryon-violating events in the BH's atmosphere, converting Higgs vacuum energy into matter under these extreme gravitational conditions. Such matter-generating regions could emerge around stellar-mass BHs, intermediate-mass BHs (IMBHs) \cite{imbh1}, or supermassive BHs (SMBHs) \cite{smbh1, smbh2, smbh3}.

The elevated rate of sphaleron transitions is sustained near the BH horizon, forming a thin atmospheric layer that acts as a site for baryogenesis. Beyond this layer, sphaleron transitions are suppressed, confining matter generation to this narrow boundary. This process could contribute to the growth of the BH’s mass and potentially affect its surrounding environment.

For IMBHs and SMBHs, the occurrence of baryon violation and the generation of ionized matter jets traveling at near-light speeds \cite{Blandford:2018iot, Jets} can create the nonequilibrium conditions required by Sakharov's criteria for baryogenesis. This dynamic could induce additional nonequilibrium effects within the system, fostering an environment conducive to sustained baryon asymmetry production.

The phenomenon of matter generation near BH horizons may help address several longstanding puzzles in cosmology and astrophysics, as discussed below.

The persistent presence of a thin, baryon-generating atmosphere surrounding a BH, though minimal in extent compared to the Schwarzschild radius, suggests ongoing matter production. This process can contribute to the growth of the BH's mass and, in the case of rotating BHs, eject matter via relativistic jets. Consequently, this mechanism could provide an alternative explanation for the rapid growth of galaxies in the early universe, as suggested by recent JWST observations \cite{jwst1,Harikane:2022rqt, Maiolino:2023bpi, Greene:2024phl}. These findings indicate that fast galaxy formation may imply the existence of primordial SMBHs, potentially reaching masses as high as $10^9 M_{\odot}$ \cite{jwstSMBH1}.

Moreover, recent JWST observations have identified SMBHs with masses in the range of $10^6 M_{\odot}$ to $10^7 M_{\odot}$
at redshifts $z \gtrsim 8$
\cite{smbhBeforeJWST1, smbhJWST1, smbhJWST2, smbhJWST3, smbhJWST4, Xiao, smbhJWST5, smbhJWST6, smbhJWST7}.
Some of these early-universe BHs appear even more massive than their host
galaxies~\cite{smbhJWST5, smbhJWST6, smbhJWST7}. This discovery challenges current models of BH and galaxy co-evolution, suggesting that SMBHs may have formed and grown at an accelerated rate during the universe's infancy, potentially indicating alternative formation mechanisms. One possible explanation for this rapid growth could be a scenario where primordial SMBH seeds gained mass predominantly through baryogenesis within their surrounding atmospheres. In this process, modifications to the Higgs potential near the BH’s horizon could significantly enhance SMBH mass growth, leading to the observed high SMBH-to-host-galaxy mass ratios.

Neutron stars (NSs) and BHs represent the final evolutionary stages of massive stars. Observations indicate that NSs have an upper mass limit between $2.0$ and $2.5~M_{\odot}$, while BHs with masses below $5~M_{\odot}$ are rarely detected \cite{mg1, mg2, mg3, mg4}. The mass range between the heaviest NSs and the lightest BHs is often referred to as the "low-mass gap." If future observations confirm the existence of this gap, it would highlight a significant challenge in our understanding of stellar evolution, potentially suggesting unconventional mechanisms in the evolution of stellar-mass BHs. The proposed mechanism of matter generation within a BH’s atmosphere could help explain how BHs with masses near the Oppenheimer-Volkoff limit can evolve into larger masses, avoiding the low-mass gap. This growth may be driven by enhanced baryon number generation processes in the extreme gravitational fields of BHs, facilitating the rapid mass increase of stellar-mass BHs.

The organization of this paper is as follows: Section \ref{sphal1} provides an overview of sphaleron processes in Schwarzschild spacetime. Section \ref{sing} discusses the necessity and challenges of modifying the classical description of BHs at horizon scales, examining potential deviations arising from strong gravitational effects. In Section \ref{higgsGrav}, we address the regularization of the Higgs potential near a BH. Within the shell assumed to surround the BH, the gravitational field is extremely strong, necessitating modifications to particle physics models, particularly adjustments to the Higgs potential. Section \ref{sphalGrav} evaluates the impact of BHs on baryon number generation rates. Section \ref{massGen} investigates enhanced baryon generation near BH horizons and its implications for early galaxy formation and the low-mass gap puzzle. Finally, we briefly conclude in Section \ref{concl}.


\section{Sphaleron process in Schwarzschild space-time} \label{sphal1}

The Lagrangian of the SM is invariant under $\mathcal{B}$-symmetry, which prohibits baryon number violation at the classical level or in any order of perturbation theory, as the baryon number current is conserved ($\partial_\nu j^\nu_{\mathcal{B}} = 0$). The lowest-dimensional, non-renormalizable $\mathcal{B}$-violating operators permitted by the SM gauge symmetries are dimension-six and thus highly suppressed. Nonetheless, within the SM framework, baryon number can be violated non-perturbatively through the EW chiral anomaly. At the quantum level, $\mathcal{B}$-symmetry is anomalous, and the baryon current is no longer conserved \cite{Adler:1969gk, Bell:1969ts, tHooft:1976rip}:
\begin{equation} \label{anomaly}
\partial_\nu j^\nu_{\mathcal B}=  \frac {n_f g^2}{64\pi^2 } \, \varepsilon^{\alpha\beta\gamma\delta}  F^a_{\alpha\beta} F^a_{\gamma\delta}~,
\end{equation}
where $n_f = 3$ denotes the number of fermion families, $F^a_{\alpha\beta}$ is the $SU(2)$ field strength tensor of the gauge field $W^a_\nu$, and $g$ is the coupling constant. The EW chiral anomaly is linked to the vacuum structure of the SM, allowing a change in the baryon number $\mathcal{B}$ (as well as the lepton number $\mathcal{L}$) \cite{tHooft:1976rip}. The same relation applies to the lepton number current, $\partial_\nu j^\nu_{\mathcal{L}}$, ensuring the conservation of $\mathcal{(B-L)}$ in the SM. However, the non-conservation of $\mathcal{(B+L)}$ is expected to have significant implications for the matter-antimatter asymmetry in the Universe \cite{Kuzmin:1985mm, Fukugita:1986hr, Shaposhnikov:1987tw}.

The change in $\mathcal{B}$-number is tied to the dynamics of transitions between non-trivial EW vacuum states. Starting at time $t = 0$, this change can be expressed as:
\begin{equation} \label{j}
\mathcal B(t) - \mathcal B(0) = \int d^4 x \, \partial_\nu j^\nu_{\mathcal B} =
\int d^3 x \, j^0_{\mathcal B} \,\bigg|_0^{t} + \int_0^t \int_S j^i \, d_i S~,
\end{equation}
where $S$ is the surface area at radius $R$. In smooth spacetime, the anomaly term \eqref{anomaly} is a generally covariant density, resulting in scalars upon integration. When covariant derivatives are well-defined, no metric tensor is needed in the integral \eqref{j}. For open, unbounded spaces, the integration can be envisioned over a spherical surface at a very large radius. If $j^i$ decreases rapidly enough at the boundary, the surface term in \eqref{j} vanishes, yielding:
\begin{equation} \label{B}
\mathcal B(t) - \mathcal B(0) =  \int d^3 x \, j^0_{\mathcal B}\,\bigg|_0^{t}
= 3\left[ n_{\rm cs} (t) - n_{\rm cs} (0)\right] ~,
\end{equation}
where
\begin{equation} \label{CS}
n_{\rm cs} (t) = \frac{g^3}{96\pi^2} \,\varepsilon^{ijk} \varepsilon^{abc} \int d^3 x\, W^a_i W^b_j W^c_k~,
\end{equation}
defines the topological Chern-Simons number in the gauge $W^a_0 = 0$. Since Chern-Simons numbers are topological, their values can be influenced by behavior at large distances. To ensure validity, it must be possible to express these scalar densities as divergences of vector densities so that, by Gauss’ theorem, their volume integral can be reduced to a surface integral. However, the regularity condition often required for applying Gauss’ and Stokes' theorems is frequently neglected in the literature, as highlighted in \cite{Lan-Lif}.

In the static case, integer values of \eqref{CS} correspond to pure gauge configurations with zero energy, residing in distinct topological vacuum states. Between these minima, the configurations have positive energy, forming an energy barrier. The energy of the saddle point configuration, known as the sphaleron, represents the height of this barrier \cite{Manton:1983nd, Klinkhamer:1984di}. To change $n_{\rm cs}$ by $\pm 1$, thereby altering the baryon number by a multiple of $n_f$, this energy barrier must be overcome. In flat Minkowski spacetime at zero temperature, the quantum tunneling rate through this barrier is exponentially suppressed, making $\mathcal{B}$-violation negligible in such conditions.

In contrast, the presence of a BH can significantly enhance the probability of sphaleron transitions \cite{DeLuca:2021oer}. The vacuum decay rate is given by:
\begin{equation} \label{rate}
\Gamma_{\rm Sph}(M_{\rm BH}) \sim \sqrt{\frac{B}{2\pi M_{\rm BH}^2}} e^{-B},
\end{equation}
where $M_{\rm BH}$ is the mass of the BH, and $B$ represents the difference between the Euclidean action of the bounce solution and that of the initial state prior to the transition. For a static sphaleron solution, it is often assumed that $B$ can be expressed as the difference between the areas at the Schwarzschild horizon and at infinity \cite{Gregory:2013hja}:
\begin{equation}
\label{B-Sph}
B \approx - \frac 14 \left( \mathcal{A}_{\rm Sch} - \mathcal{A}_\infty \right) \approx E_{\rm Sph}(M_{\rm BH})M_{\rm BH}~,
\end{equation}
where $E_{\rm Sph}(M_{\rm BH})$ is the sphaleron energy in the presence of a BH. Similar to \eqref{rate}expression have been proposed for the decay rate of the false vacuum \cite{Coleman:1980aw}, BH nucleation \cite{Mellor:1989wc, Dowker:1993bt}, and open universe nucleation \cite{Hawking:1998bn}. However, the only context where estimates like \eqref{rate} are rigorously justified is in flat spacetime \cite{Coleman:1977}. The energy $E_{\rm Sph}(M_{\rm BH})$ may also vary depending on the proximity of the sphaleron transition processes to the BH’s horizon.


\section{Horizon singularities} \label{sing}

In this paper, we examine the Higgs vacuum within the gravitational field of a non-rotating spherical body, described by the Schwarzschild metric:
\begin{equation} \label{Schwarzschild}
ds^2 = \frac{(r - 2M)}{r}\, dt^2 - \frac{r}{(r - 2M)}\, dr^2 - r^2 d\theta^2 - r^2 \sin^2 \theta\, d\phi^2~.
\end{equation}
This metric is typically assumed to remain valid even 'inside' the Schwarzschild BH, i.e., for $r < 2M$. Here, the 'mass' $M$ is not localized at a specific point (since the stress-energy-momentum tensor is zero), but serves as a useful parameter for describing a BH that has not yet fully formed from the perspective of a distant observer. After complete collapse, all matter that formed the BH is expected to vanish into the central singularity at $r = 0$, leaving behind only the vacuum gravitational field \cite{BH-1, BH-2, BH-3, Wald, gravMTW}.

The Schwarzschild solution also features a horizon singularity at $r = 2M$, which results in divisions by zero or multiplications by infinity in certain geometrical quantities. However, his singularity at $r = 2M$ is a coordinate singularity that can be eliminated by switching to more suitable coordinates. In contrast, the point $r = 0$ corresponds to a true physical singularity, evident in coordinate-independent quantities such as the Kretschmann invariant:
\begin{equation} \label{Kretschmann}
{\mathcal R}^{\alpha\beta\gamma\delta}{\mathcal R}_{\alpha\beta\gamma\delta} = \frac {48 M^2}{r^6} ~.
\end{equation}
For sufficiently large BHs, the Kretschmann invariant \eqref{Kretschmann} can be made arbitrarily small as $r \to 2M$, suggesting that an astrophysical BH cannot exhibit significant curvature at the horizon.

The concept of a coordinate singularity is the primary motivation for using alternative radial variables instead of $r$, ensuring that the horizon singularity in the Schwarzschild metric \eqref{Schwarzschild} is 'eliminated'. This allows classical particles to traverse the horizon and reach the central naked singularity at $r = 0$ without obstruction. A key element of all singular coordinate transformations of the Schwarzschild metric is the Regge-Wheeler tortoise coordinate. However, this coordinate does not belong to the $C^2$ class of admissible functions, which means the singular transformations (such as those introduced by Kruskal-Szekeres, Eddington-Finkelstein, Lemaitre, or Gullstrand-Painlevé) generate delta functions in the second derivatives (see, for example, \cite{gravMTW, meG4} for details). This results in transformed metric tensors that are not differentiable at the horizon, placing them in the $C^0$ class, which modifies the Einstein equations by introducing spurious delta-function sources at the horizon.

To address the issues posed by singularities, distributional approaches for metrics can be utilized, regularizing them through sequences of smooth functions \cite{Colombeau-1, Colombeau-2, Gel-Sch, Pantoja:1997zt, Heinzle:2001bk, Foukzon}. The claim of the Kretschmann invariant \eqref{Kretschmann} remaining finite at $r = 2M$ relies on the mathematically questionable assumption of mutual cancellations of delta-function-like divergences.

Consider the regularized Schwarzschild radial variable \cite{Foukzon, Foukzon1}:
\begin{equation} \label{r+epsilon}
r \quad \to \quad r_\epsilon = 2M + \sqrt{\left( r - 2M \right)^2 + \epsilon^2}~,
\end{equation}
where $\epsilon$ is an infinitesimal length parameter. In this coordinate system, the Ricci scalar exhibits singular behavior at the horizon $r \to 2M$:
\begin{equation} \label{Ricci}
{\mathcal R}_\epsilon \big|_{\epsilon \to 0} \approx \frac {\epsilon^2}{2M\left[\left( r - 2M \right)^2 + \epsilon^2\right]^{3/2}}\bigg|_{\epsilon \to 0} \sim \delta (r - 2M) \quad \to \quad \infty~.
\end{equation}
For the generalized Kretschmann scalar, we also find a singular expression:
\begin{equation} \label{Kretschmann-2}
{\mathcal R}^{\alpha\beta\gamma\delta}{\mathcal R}_{\alpha\beta\gamma\delta} \big|_{\epsilon \to 0} \approx \frac {48 M^2}{r^6} + \frac {\epsilon^4}{4M^4\left[\left( r - 2M\right)^2 + \epsilon^2\right]^3}\bigg|_{\epsilon \to 0} \quad \to \quad \infty~.
\end{equation}
Thus, the apparent smallness of the Kretschmann scalar \eqref{Kretschmann} does not indicate that the curvature is small at the BH horizon. This is because three out of the six non-zero independent components of the mixed Riemann tensor for the Schwarzschild metric diverge as $r \to 2M$.

The singular behavior of the Riemann and Ricci tensors at the horizon suggests the necessity of assuming that a physical BH is surrounded by a thin spherical shell, located between the Schwarzschild radius and the photon sphere, specifically in the range $2M \leq r \leq 3M$. In this region, geometric quantities are regularized, enabling a smooth transition between the external vacuum metric \eqref{Schwarzschild} and an internal nonsingular metric. This internal metric describes a real, spherically symmetric massive object governed by a 'quantum' equation of state.

The need to modify the classical description of BHs at horizon scales arises from the incompatibility of general relativity with the principles of quantum mechanics \cite{Modif-1, Modif-3}. Nevertheless, much of the scientific community has been hesitant to abandon the traditional concept of BH singularities, a notion that has been widely accepted for nearly a century. However, this perspective may be shifting in light of Roy Kerr's recent paper \cite{Kerr:2023rpn}, which argues that the interior of a BH contains a physical object rather than empty space with a central singularity. It is expected that quantum gravitational effects dominate within a BH, potentially enabling a regular description of spacetime in a quantum framework.

Observationally, the interior of a BH remains inaccessible \cite{Cardoso:2019rvt}, but the possibility of BHs without singularities has been suggested \cite{Cardoso:2019rvt, Ansoldi:2008jw}. If BHs are indeed 'normal' astrophysical objects, their event horizons would enclose an impenetrable sphere of ultra-dense 'quantum' matter \cite{Malafarina:2017csn}, possibly leading to observable phenomena such as wave reflections from the event horizon \cite{meG1, meG2, meG3}. This emerging view reconciles two previously conflicting perspectives on the BH horizon. One perspective holds that the energy density at the horizon is small \cite{Howard:1984qp}, while the other proposes the existence of structures such as a 'brick wall' \cite{tHooft:1984kcu}, 'fuzzball' \cite{Mathur:2005zp}, or 'firewall' \cite{Modif-2}, which would significantly alter the surrounding geometry.

It is noteworthy that, like many modern physicists, both Einstein and Schwarzschild believed that matter could not cross the event horizon \cite{Einstein, Schwarz}.

In particle physics, the Schwarzschild solution is often more conveniently expressed in isotropic coordinates, as presented in \cite{Wald}. The metric in these coordinates is given by:
\begin{equation} \label{metric}
ds^2 = N^2(R)\, dt^2 - \psi^4(R) \left( dR^2 + R^2 d\theta^2 + R^2 \sin^2 \theta\, d\phi^2 \right)~,
\end{equation}
where the metric components are determined by the lapse function:
\begin{equation} \label{N}
N(R) = \frac{1 - M/2R}{1 + M/2R}~, \qquad \qquad (0 \leq N(R) \leq 1)
\end{equation}
and the conformal factor:
\begin{equation} \label{psi}
\psi(R) = 1 + \frac{M}{2R}~, \qquad \qquad (1 \leq \psi(R) \leq 2)~.
\end{equation}
As $R$ approaches infinity, corresponding to $r \rightarrow \infty$, the isotropic radial coordinate $R$ can be related to the Schwarzschild coordinate $r$ by the following expression:
\begin{equation} \label{2R}
r = R \left( 1 + \frac{M}{2R} \right)^2~, \qquad 2R = r - M + \sqrt{r^2 - 2Mr}~.
\end{equation}
The metric \eqref{metric} is particularly advantageous for studying particle physics in the gravitational field of a BH because it features a conformally Euclidean spatial part and avoids singularities at the horizon. Additionally, it facilitates regularization, such as modifying the lapse function as proposed in \cite{Foukzon, Foukzon1}:
\begin{equation}
N_\epsilon(R) = \frac{\sqrt{(R - M/2)^2 + \epsilon^2}}{R + M/2}~.
\end{equation}

In isotropic coordinates, the Schwarzschild horizon and the photon sphere are located at $R = M/2$ and $R = (2 + \sqrt{3})M/2$, respectively. Although the radial part of the isotropic metric \eqref{metric} is regular at $R = M/2$, it still leads to physical singularities when applied to matter field equations. This is due to the determinant of the metric:
\begin{equation} \label{det}
\sqrt{-g} = N\, \psi^6\, R^2 \sin\theta = \left(1 - \frac{M}{2R}\right)\left(1 + \frac{M}{2R}\right)^5 R^2 \sin\theta~,
\end{equation}
which vanishes at the horizon.  The lapse function $N(R)$ approaches zero at $R = M/2$, implying that isotropic coordinates cannot extend through the horizon. Therefore, these coordinates are valid only for the region outside the Schwarzschild radius, where $r > 2M$.

In scenarios where classical particles cannot cross the horizon \cite{meG1, meG2, meG3}, isotropic coordinates (remaining nonsingular at the horizon) offer a more accurate description of the background geometry. In contrast, the singularities introduced by coordinate transformations in the Schwarzschild gauge obscure the nonsingular nature of the metric.


\section{Higgs vacuum at the horizon} \label{higgsGrav}

BHs can be modeled as being surrounded by a narrow atmospheric layer of thickness $\epsilon$, which is generated by a delta-like source at the Schwarzschild horizon and influenced by the intense gravitational field near the horizon. This setup assumes that the wavefunctions of matter fields vanish both at $r = 2M$ and at $r = 2M + \epsilon$, in accordance with the brick-wall model proposed by 't Hooft \cite{tHooft:1984kcu}. Within this shell, the gravitational potential is expected to be extremely strong, leading to distinctive effects in particle physics models.

As an example, we consider the regularization of the Higgs potential:
\begin{equation} \label{U_0}
U_0  = \frac {M_h^2}{2}h^2 + \lambda v h^3 + \frac {\lambda}{4} h^4~,
\end{equation}
where $\lambda \approx 0.13$ is the self-coupling constant, $v$ is the vacuum expectation value (VEV), and $M_h = \sqrt{2\lambda} v$ represents the physical Higgs scalar mass. This low-energy potential has a minimum at $U_0 = 0$ when $h = 0$ and $v = 246$~GeV, corresponding to the EW vacuum, which is stable at the tree level. In a strong gravitational potential, the Higgs potential acquires radiative corrections, requiring the introduction of a generalized regularization scheme \cite{Chitishvili:2021jrx}.

In quantum field theory in Minkowski spacetime, standard regularization arises from specific integrals \cite{Col-Wei}. For example, for a scalar particle of mass $m$, we have \cite{Peskin_Book, Schwartz:2013pla}:
\begin{equation} \label{int}
\int \frac {dEd^3p}{(2\pi)^4} \frac {1}{\left(E^2 - p^2 - m^2 + i\varepsilon\right)^2} \to - \frac {i}{16\pi^2} \ln \frac{m^2}{\Lambda^2}~,
\end{equation}
where $\Lambda$ is the regularization scale. This implies that an observable differs from its value at higher energy scales $\sim \Lambda$ by a logarithmic term. When quantized, the Higgs field potential \eqref{U_0} is modified to:
\begin{equation} \label{U_SM}
U_{\rm eff}(\Lambda) = U_0  + U_1(\Lambda)~,
\end{equation}
with the radiative corrections,
\begin{equation} \label{U_1}
U_1(\Lambda) = \sum_i \frac {n_i}{64\pi^2} M_i^4 \left[\ln \left(\frac {M^2_i}{\Lambda^2}\right) - C_i\right]~,
\end{equation}
where the sum runs over particle species, $n_i$ counts the degrees of freedom (with a minus sign for fermions), $C_i$ are constants. The field-dependent mass squared of the $i$-th species is:
\begin{equation}
M^2_i (h) = \Lambda^2_i + g_i h^2~,
\end{equation}
where $g_i$ are the coupling constants \cite{Schwartz:2013pla}.

In \eqref{U_1}, the regularization scale can be chosen as the VEV of the Higgs field, $\Lambda \simeq v$, since this is the scale at which the running SM parameters still align with experimentally observed values and the Higgs potential is well approximated by its classical form. At higher energies, $h \simeq 10 v$, the logarithmic terms in \eqref{U_1} become positive and larger than the constants $C_i$, so the effective potential \eqref{U_SM} is well approximated by:
\begin{equation} \label{U1}
U_{\rm eff}(h)  \approx \left[\frac {\lambda}{4} + \frac {3}{64\pi^2} \left( 3\lambda^2 - Y_t^4 \right) \ln \frac {h^2}{v^2}\right] h^4~,
\end{equation}
where $Y_t = \sqrt 2 M_t / v \approx 0.99$ is the $t$-quark Yukawa coupling. Due to the minus sign in front of the $t$-quark term in \eqref{U1}, the one-loop logarithmic correction becomes negative, meaning the SM effective potential $U_{\rm eff}$ develops a local minimum. However, due to the large denominator, $64\pi^2$, the second term in \eqref{U1} is smaller than $\lambda / 4$, so the potential \eqref{U1} remains positive. Thus, in Minkowski space calculations, the minimum of $U_{\rm eff}$ is still higher than the EW vacuum value $U_0 = 0$. As a result, the bounce action \eqref{B-Sph}, which is primarily determined by the Higgs potential $U_{\rm eff}$, remains large, and the probability of baryon number violation \eqref{rate} is exponentially suppressed.

In strong gravitational fields, the dispersion relations of particles and regularized integrals of the type \eqref{int} must be modified \cite{Chitishvili:2021jrx}. The singular behavior of the Schwarzschild metric and related geometric quantities renders the mathematical meaning of certain particle physics concepts near a BH horizon imprecise. For instance, gauge transformations of the Higgs field kinetic term, $g^{\mu\nu}(D_\mu H)^\dag D_\nu H$, on the background \eqref{metric} become ill-defined at the horizon, where $g^{00} \to \infty$. However, the failure of particle physics models in strong gravitational fields does not necessarily imply the breakdown of the geometry itself. Instead, the metric should be regularized around the horizon.

The regularized Ricci scalar \eqref{Ricci} contains a $\delta$-like term. This will modify the bounce action \eqref{B-Sph} by the following term:
\begin{equation} \label{B'}
\begin{split}
\int d^4x \sqrt{-g_\epsilon} {\mathcal R}_\epsilon \quad \to \quad \int_{2 M_{\rm BH}}^{2 M_{\rm BH} + \epsilon} dr \frac {r_\epsilon^2}{2 M_{\rm BH} \left[ \left( r - 2 M_{\rm BH} \right)^2 + \epsilon^2 \right]^{3/2}} \approx \\
\approx \frac {2 M_{\rm BH}}{\sqrt {\left( r - 2 M_{\rm BH} \right)^2 + \epsilon^2}}\bigg|_{2 M_{\rm BH}}^{2 M_{\rm BH} + \epsilon} \approx \frac {2 M_{\rm BH}}{\epsilon}~,
\end{split}
\end{equation}
which contributes significantly to the sphaleron energy for any realistic particle physics scale.

In isotropic coordinates, the entire radiative corrections term in \eqref{U_1} will be modified by the universal factor \cite{Chitishvili:2021jrx}:
\begin{equation} \label{hMod2}
U_1 \to \frac {\psi^3}{N} \, \frac {3}{64\pi^2} \left( 3\lambda^2 - Y_t^4 \right) \ln \frac {h^2}{v^2} \, h^4~.
\end{equation}
Sufficiently close to the BH horizon, the lapse function $N (R) \to 0$, and the negative one-loop correction term in \eqref{hMod2} will reduce the value of the Higgs effective potential \eqref{U1}, potentially even pushing it into the negative region. As a result, the bounce action $B$ approaches zero, implying that the baryon number violation rate \eqref{rate} can increase. Therefore, in this scenario, baryon number violation through sphaleron transitions near the BH horizon can occur at a much faster rate compared to flat Minkowski spacetime.


\section{Dynamic Chern-Simons} \label{sphalGrav}

To estimate the influence of BHs on the change in in the baryon number $\mathcal B$ (as defined in \eqref{B}), a model with the static Higgs doublet field $H$ and an $SU(2)_L$ gauge field $W^a_\nu$ was studied \cite{DeLuca:2021oer}. The $U(1)_Y$ gauge field is neglected in this context, as its contribution to the sphaleron energy is found to be negligible. The following spherically symmetric sphaleron ansatz is used \cite{Klinkhamer:1984di}:
\begin{equation} \label{ansatz}
\begin{split}
H(r) &= \frac{v}{\sqrt{2}}\, h(r)\, U
\begin{pmatrix}
0\\1
\end{pmatrix}, \\
\frac {\sigma^a}{2}W^a_i(r) &= - \frac {i}{g}\, f(r)\, \partial_i U (U)^{-1}~,
\end{split}
\end{equation}
where $g$ is the $SU(2)_L$ coupling constant, and the matrix $U = i \, \mathcal{N}_i \sigma^i$ represents an element of $SU(2)$, with $\mathcal{N}^i$ denoting the unit normal vector to the two-sphere boundary and $\sigma^i$ the Pauli matrices. In the Schwarzschild metric \eqref{Schwarzschild}, the unit normal vector to the two-sphere boundary,
\begin{equation} \label{N-i}
\mathcal{N}^i = \sqrt {\frac {r}{r-2M}} \, \frac {x^i}{r}~,
\end{equation}
becomes ill-defined at the horizon. To resolve this issue, a regularized radial variable $r_\epsilon$ defined in \eqref{r+epsilon} is introduced.

In \eqref{ansatz}, the dimensionless radial functions of the Higgs field $h(r)$ and the isospin gauge field $f(r)$ must satisfy the following boundary conditions:
\begin{equation} \label{h,f}
\begin{split}
h(r \to 2 M_{\rm BH}) &= f(r \to 2 M_{\rm BH}) \to 0~ , \\
h(r \to \infty) &= f(r \to \infty) \to 1~.
\end{split}
\end{equation}
The first boundary condition ensures the absence of singularities at the Schwarzschild horizon, while the second ensures that the ansatz asymptotically matches the expected form of the fields at infinity.

For the sphaleron ansatz in \eqref{ansatz}, the Chern-Simons current in \eqref{j} takes the following form:
\begin{equation}
j^\mu = \frac{2}{3} f^3(r) \, \varepsilon^{\mu\nu\alpha\beta} \, \text{Tr} \left[ U \partial_\nu (U)^{-1} U \partial_\alpha (U)^{-1} U \partial_\beta (U)^{-1} \right]~.
\end{equation}
To express $U$ in terms of the unit vector and the parameter along the loop in configuration space $\mu$ \cite{Manton:1983nd}, we write:
\begin{equation}
U = e^{i \, \mathcal{N}_i \sigma^i \, \mu f(r)} = \mathbf{I} \cos \left( \mu f(r) \right) + i \, \mathcal{N}_i \sigma^i \sin \left( \mu f(r) \right)~.
\end{equation}
It is important to note that the static solution only exists at an extremum of the energy functional of the model, which occurs at half-integer multiples of $\mu/\pi$. Away from these values, no static solution exists.

Assuming for simplicity that the Chern-Simons current vanishes at $t = 0$ (i.e., $j^\nu(t = 0) = 0$ or $n_{cs}(0) = 0$ as defined in \eqref{B}), and using the boundary conditions \eqref{h,f}, the Chern-Simons number at time $t$ is given by \cite{Tye:2016pxi}:
\begin{equation} \label{n-cs}
\begin{split}
n_{cs}(t) &= \int d^3x \, j^0 \\
&= \frac{1}{2\pi} \left[ 2\mu f(r_\epsilon) - \sqrt{\frac{r_\epsilon}{r_\epsilon - 2M}} \sin 2\mu f(r_\epsilon) \right]_{r = 2 M_{\rm BH}}^{r = \infty} \\
&= \frac{1}{2\pi} \left( 2\mu - \sin 2\mu \right) + \frac{\sqrt{2 M_{\rm BH}}}{2\pi} \frac{\sin 2\mu f(2 M_{\rm BH} + \epsilon)}{\sqrt{\epsilon}} \bigg|_{\epsilon \to 0}~.
\end{split}
\end{equation}
If the sphaleron profile function at $r = 2 M_{\rm BH} + \epsilon$ has the asymptotic behavior:
\begin{equation}
f(2 M_{\rm BH} + \epsilon) \big|_{\epsilon \to 0} \sim \epsilon^k~,
\end{equation}
where the parameter $k \le 1/2$, then the second term in \eqref{n-cs} becomes very large. This indicates a rapid increase of the $\mathcal{B}$-number \eqref{B} near the BH horizon.

In 't Hooft's brick wall model \cite{tHooft:1984kcu}, the wall thickness, represented by the parameter $\epsilon$, serves as a cutoff distance from the event horizon of a BH.
This distance is not a precisely measurable physical quantity; rather, it is an exceedingly small value compared to the Schwarzschild radius $R_{\rm S}$ of the BH, typically of the order of $10^{-n} R_{\rm S}$, where $n$ is a large number. The introduction of this cutoff establishes a boundary layer where quantum effects can be explored without directly interacting with the singularity at the horizon. While the specific choice of $\epsilon$ is somewhat arbitrary, it is kept sufficiently small to have minimal impact on observable physics at larger scales \cite{Ghosh:2009xg, brick1, brick2}. In the context of modifications to the Higgs potential, this atmospheric thickness could correspond to energy levels at which EW symmetry is restored, thus providing a theoretical framework for sphaleron transitions near the BH horizon.


\section{Early galaxies growth and the low-mass gap} \label{massGen}

A significant challenge in astrophysics is the observed tight correlations between BH masses and their host galaxy properties (such as bulge mass and stellar velocity dispersion), which strongly suggest that the growth of galaxies is linked to their central BHs (see the recent reviews~\cite{Zh-Ho}). Additionally, understanding the nature of the link between jets from BHs and the accompanying accretion disks is problematic \cite{Ricarte:2023owr}. Observations of astrophysical objects with slow spin rates and a lack of accretion material, such as the SMBH of our galaxy \cite{EHTC}, suggest that jets are neither rotation nor accretion powered.

Difficulties also arise with the standard assumption that massive BHs, forming from lower-mass BH 'seeds', gain most of their mass via accretion, modeled in various ways (see reviews \cite{Inayoshi:2019fun, Volonteri:2021sfo}). The co-evolution of SMBHs and their host galaxies has been extensively investigated using different approaches, including semi-analytical models based on merger trees \cite{tree01, tree1, tree2, tree3, tree4, tree5}, large-scale simulations incorporating N-body dynamics and hydrodynamics \cite{hydro1, hydro2, hydro3}, and frameworks based on the extended Press-Schechter formalism \cite{eps}. Recent findings from the JWST have identified a significant population of high-redshift galaxies hosting massive central BHs \cite{jwst1, Harikane:2022rqt, Maiolino:2023bpi, Greene:2024phl}. Surprisingly, these BHs appear to be significantly more massive compared to their galaxy hosts-by, an order of magnitude greater than what is observed in local galaxies \cite{Pacucci:2023oci, Juodzbalis}. Moreover, some of these early-universe BHs exhibit masses that even surpass those of their host galaxies \cite{smbhJWST5, smbhJWST6, smbhJWST7}. Such discrepancies underscore the potential existence of unexplored mechanisms driving the accelerated growth of BHs and galaxies, challenging conventional models of their evolution.

An enhanced sphaleron transition rate within the BH atmosphere offers a novel mechanism for matter generation, potentially explaining both early galaxy formation and the low-mass gap puzzle. We demonstrate that this mechanism can account for both phenomena under similar microscopic conditions, despite the vast difference in BH masses involved, differing by many orders of magnitude. This shared process, wherein extreme gravitational conditions modify the sphaleron transition rate and the multiplicity of baryon production, provides a unified approach to understanding both the rapid growth of SMBHs and the scarcity of BHs within certain mass ranges.

The generation of baryons can be estimated by calculating the number of sphaleron transitions within a specific volume, with each transition contributing a set number of baryon units, where the asymmetry between matter and antimatter is driven by CP violation \cite{bProd1}. A simple estimate of the total mass of baryons generated by a BH with mass $M_{\rm BH}$ over a time period $t_{\rm gen}$ can be presented as follows:
\begin{equation} \label{genBar1}
M_{\rm gen}(t_{\rm gen})\approx m_b \, n_{\rm cs}(M_{\rm BH}) \, V_{d_{\rm H}}(M_{\rm BH}) \, \langle\Gamma_{\rm sph}(M_{\rm BH},d_{\rm H})\rangle \, \delta_{\rm CP} \, t_{\rm gen} ~.
\end{equation}
Here, $m_b$ is the mass of a single baryon unit, approximated as the proton mass. The factor $n_{\rm cs}(M_{\rm BH})$ represents the number of baryon units produced per sphaleron transition within the volume $V_{d_{\rm H}}(M_{\rm BH})$ of the layer where baryon production occurs. The term $\langle \Gamma_{\rm sph}(M_{\rm BH}, d_{\rm H}) \rangle$ denotes the average sphaleron transition rate over a spherical layer of dimensionless width $d_{\rm H} = \epsilon / R_{\rm S}$, where $R_{\rm S} = 2GM_{\rm BH}$ is the Schwarzschild radius of the BH. Finally, $\delta_{\rm CP}$ accounts for the CP violation involved in the process.

Equation \eqref{genBar1} holds as long as the time $t_{\rm gen}$ is short enough to keep $M_{\rm gen}$ comparable to $M_{\rm BH}$. This condition allows us to estimate the potential of this mechanism to explain early galaxy formation and the low-mass gap, as discussed below.

We express the sphaleron rate within the BH atmosphere as:
\begin{equation} \label{PhTrR1}
{\Gamma_{\rm sph} (M_{\rm BH}},d_{\rm H}) = {\cal M}^4 e^{-B(M_{\rm BH},d_{\rm H})}~,
\end{equation}
where $\mathcal{M}$ has dimensions of energy, and $B(M_{\rm BH}, d_{\rm H})$ is an effective action that depends on the mass of the BH, $M_{\rm BH}$, and the dimensionless distance from the horizon $d_{\rm H}$, as suggested in previous studies \cite{Chitishvili:2021jrx}.

Building on the results from Section \ref{higgsGrav}, which indicate that EW symmetry may be restored within the BH atmosphere, we replace $\mathcal{M}^4$ with the sphaleron rate in the symmetric EW phase, as computed in \cite{symRate1, symRate2, symRate3}. For an approximate estimate, this rate can be expressed as:
\begin{equation} \label{sphalSym1}
{\cal M}^4 \simeq \kappa \, \alpha_W^5 \,E_{\rm EW}^4 ~,
\end{equation}
where $E_{\rm EW} \simeq 100 \, \text{GeV}$ is the energy scale at which EW symmetry is restored. This scale corresponds to the conditions where symmetry is assumed to be restored near the BH horizon, analogous to the thermal conditions during the EW epoch in the early universe.

This approximation allows us to compare processes such as baryogenesis and sphaleron transitions in BH atmospheres, especially at distances $\epsilon \lesssim 1/T_{\rm EW}$. We use this distance as an estimate for the width of the BH atmosphere. The constants in this expression are $\alpha_W = g^2/4\pi$ and $\kappa \approx 20$, determined from lattice measurements \cite{Tew1}.

For estimation purposes, following \cite{Chitishvili:2021jrx}, the distance dependence of the effective action can be modeled as:
\begin{equation} \label{effAct1}
B(M_{\rm BH},d_{\rm H}) = B_{1}(M_{\rm BH}) \, d_{\rm H}^{b}~,
\end{equation}
where $B_1(M_{\rm BH})$ is a factor that depends on the BH mass, analogous to the term appearing in \eqref{B-Sph}. The parameter $b > 0$ is an exponent that reflects the gravitational corrections to the Higgs potential, allowing the effective action to vary with distance from the BH horizon \cite{Chitishvili:2021jrx}.

Thus, the closer the sphaleron transition site is to the horizon, the higher the transition probability. As we move away from the horizon, the exponential suppression in \eqref{PhTrR1} increases. The term $\langle \Gamma_{\rm sph}(M_{\rm BH}, d_{\rm H}) \rangle$ represents the overall sphaleron rate, effectively averaging the action $B(M_{\rm BH}, d_{\rm H})$ over the entire width of the layer $d_{\rm H}$.

For further numerical estimates, the formula \eqref{genBar1} can be rewritten as:
\begin{equation} \label{genBar2}
M_{\rm gen}(t_{\rm gen})\approx 2\times 10^{5}M_{\odot}\left(\frac{\cal M}{\rm GeV}\right)^4 \langle e^{-B(M_{\rm BH},d_{\rm H})}\rangle \, n_{\rm cs}(M_{\rm BH}) \left( \frac{t_{\rm gen}}{\rm yr}\right)\left[\frac{V_{d_{\rm H}}(M_{\rm BH})}{\rm km^{3}}\right] ~,
\end{equation}
where $\delta_{\rm CP} \simeq 10^{-25}$ accounts for CP violation within the SM \cite{CPinSM1}.

The number of generated baryon units can be estimated from \eqref{n-cs} by setting:
\begin{equation} \label{fcs1}
f(2 GM_{\rm BH}+\epsilon)\approx\left(\frac{\epsilon}{2GM_{\rm BH}}\right)^k~,
\end{equation}
yielding:
\begin{equation} \label{nb1}
n_{\rm cs}(M_{\rm BH})\simeq n_{\rm\mu cs}+\frac{\sqrt{2}\mu}{\pi} \, d_{\rm H}^{(k-1/2)} ~,
\end{equation}
where $n_{\rm\mu cs}$ is a constant of order unity. For distances $d_{\rm H} \ll 1$, if $k \lesssim 1/2$, the second term dominates, implying that sphaleron transitions generate a substantial number of baryon units. Conversely, for $k > 1/2$, each sphaleron transition results in an increment of approximately one baryon unit.

The volume of the baryon production layer can be approximated as:
\begin{equation} \label{vol1}
V_{d_{\rm H}}(M_{\rm sBH}) \approx 4\pi R_{\rm S}^3 \, d_{\rm H} = 32\pi G^3 M_{\rm BH}^3 \, d_{\rm H} ~,
\end{equation}
where $R_{\rm S} = 2GM_{\rm BH}$ is the Schwarzschild radius.

We proceed with our estimates in light of the recent JWST discoveries regarding early galaxy formation \cite{jwst1}. JWST observations have revealed massive, compact galaxies, each with a mass around $10^{10} M_{\odot}$, and some reaching up to $10^{11} M_{\odot}$. These galaxies, densely packed with stars, appear to have formed only 700 million years after the Big Bang, much earlier and more rapidly than previously expected. This rapid formation suggests the possible presence of SMBHs with masses reaching up to $10^9 M_{\odot}$ \cite{jwstSMBH1}.

Here, we attribute the mass generation of the early galaxies observed by JWST to sphaleron transitions near the SMBH horizon. If the mass generated in the proximity of a SMBH with mass $10^9 M_{\odot}$, as estimated from \eqref{genBar2}, reaches $10^{10} M_{\odot}$ within a timescale of $\lesssim 1$ billion years, then the following condition must hold:
\begin{equation} \label{GenBySMBH1}
B(M_{\rm BH} = 10^{9}M_{\odot},d_{\rm H})\approx 50-70k\ ,
\end{equation}
where we consistently assume $k < 1/2$ throughout the analysis below.
This result is obtained by setting ${\cal M}\approx 10$~GeV, as inferred from \eqref{sphalSym1}, with $\epsilon\simeq 10^{-16}$~cm ($d_{\rm H}\simeq 10^{-30}$). This length scale corresponds to the energy level $E_{\rm EW}\approx 100\ {\rm GeV}$, matching the threshold at which EW symmetry is restored. By comparing the effective action \eqref{effAct1}, expressed as:
\begin{equation} \label{GenBySMBH2}
B(M_{\rm BH},d_{\rm H})\approx 8\pi G \, E_{\rm sph} \, M_{\rm BH} \, d_{\rm H} ~,
\end{equation}
with (\ref{GenBySMBH1}), we arrive at the estimate:
\begin{equation} \label{GenBySMBH3}
\left(\frac{E_{\rm sph}}{M_{\rm Pl}}\right)_{10^{9}M_{\odot}}\approx 10^{-17}(1-1.4k) ~,
\end{equation}
indicating that the sphaleron energy may correspond to the EW scale, $E_{\rm sph}\approx 100$~GeV.

A similar estimate aligns with recent observations from the JWST, which reveal SMBHs in the mass range of approximately $10^6 M_{\odot}$ to $10^7 M_{\odot}$ at redshifts $z \gtrsim 8$ \cite{smbhBeforeJWST1, smbhJWST1, smbhJWST2, smbhJWST3, smbhJWST4, smbhJWST5, smbhJWST6, smbhJWST7}. Remarkably, some of these early-universe BHs appear even more massive than their host galaxies \cite{smbhJWST5, smbhJWST6, smbhJWST7}. For these SMBHs, the condition becomes:
\begin{equation} \label{GenBySMBH4}
B(M_{\rm BH}=10^{6}M_{\odot},d_{\rm H})\approx 54-62k ~.
\end{equation}
This estimate corresponds to the mass generated near the horizon of a SMBH with mass $10^6 M_{\odot}$, which reaches about $10^7 M_{\odot}$ within a timescale of less than 0.2 billion years. Here, we employ the scale $d_{\text{H}} \approx 10^{-27}$, calculated with $\epsilon \approx 10^{-16}$~cm. Thus, for the sphaleron process, we estimate:
\begin{equation} \label{GenBySMBH5}
\left(\frac{E_{\rm sph}}{M_{\rm Pl}}\right)_{10^{6}M_{\odot}}\approx 10^{-17}(2.2-2.5k) ~.
\end{equation}
This suggests again a sphaleron energy scale consistent with the EW scale, potentially providing insights into the rapid growth of SMBHs and their role in early-universe galaxy formation.

The mechanism of matter generation described here can also explain the origin of the low-mass BH gap. Rather than altering our understanding of stellar evolution or BH formation, we propose that BHs formed within the low-mass gap proceed to gain mass by producing matter near their horizons and gradually accreting it. This gradual accumulation leads to an increase in their mass, eventually moving them into the upper range of the gap, around $5 M_{\odot}$.

By applying formula \eqref{genBar2} to estimate the parameters needed for a low-mass gap BH, initially around $3 \, M_{\odot}$, to evolve into a BH with a mass of $5 \, M_{\odot}$ over a typical galactic age of 10 Gyr, one finds the condition:
\begin{equation} \label{GenBySMBH6}
B(M_{\rm BH}=3M_{\odot},d_{\rm H})\approx 40-22k ~,
\end{equation}
where we used $d_{\text{H}} \approx 10^{-22}$, again calculated with $\epsilon \approx 10^{-16}$~cm. Therefore, in this case, the sphaleron energy is given by:
\begin{equation} \label{GenBySMBH7}
\left(\frac{E_{\rm sph}}{M_{\rm Pl}}\right)_{3M_{\odot}}\approx 10^{-17}(4-2.2k)~,
\end{equation}
which is again aligned with the EW energy scale.


\section{Conclusions} \label{concl}

In this work, we have proposed that modifications to the Higgs potential within the narrow atmospheric layer near the event horizon of a BH can substantially enhance the rate of sphaleron transitions, effectively transforming the Chern-Simons number into a dynamic variable. Unlike the usual suppression of sphaleron transitions under low-temperature conditions, the gravitational effects near the BH horizon facilitate these transitions, eliminating the suppression and allowing them to proceed more readily. Furthermore, each transition near the horizon is likely to generate a significantly higher baryon number compared to what would be expected in a typical Minkowski spacetime, thus amplifying baryon number violation around the BH.

Our findings demonstrate that the enhanced sphaleron transition rate, influenced by gravitational corrections near the event
horizon, plays a critical role in the generation of matter around SMBHs. This process may contribute to the rapid mass growth
of these objects and the formation of their host galaxies.
The connection between BH mass and the rate of baryon production provides critical insights into
the timescales and conditions required for the formation of early SMBHs. These findings are consistent with
recent high-redshift observations captured by the JWST, which highlight rapid BH growth during the early
galaxies formation stage.
Additionally, this mechanism offers a novel
explanation for the low-mass gap in the black hole mass spectrum, potentially accounting for the
observed scarcity of black holes near the Oppenheimer-Volkoff limit.

Overall, our results highlight the importance of sphaleron transitions in early galaxy evolution, suggesting that they could be a key factor in the interplay between BH growth and baryon generation. Future observational and theoretical research, particularly through gravitational wave astronomy and further high-redshift surveys, will be essential in testing the predictions of this model. Such investigations will help advance our understanding of the complex relationship between BH growth, galaxy formation, and the dynamics of the early universe.



\begin{thebibliography}{999}

\bibitem{Riotto:1999yt} A.~Riotto and M.~Trodden,
``Recent progress in baryogenesis,''
Ann. Rev. Nucl. Part. Sci. \textbf{49} (1999) 35,
doi: 10.1146/annurev.nucl.49.1.35
[arXiv: hep-ph/9901362].

\bibitem{Banks:2010zn} T.~Banks and N.~Seiberg,
``Symmetries and strings in field theory and gravity,''
Phys. Rev. D \textbf{83} (2011) 084019,
doi: 10.1103/PhysRevD.83.084019
[arXiv: 1011.5120 [hep-th]].

\bibitem{Antunes:2019phe} V.~Antunes, I.~Bediaga and M.~Novello,
``Gravitational baryogenesis without CPT violation,''
JCAP \textbf{10} (2019) 076,
doi: 10.1088/1475-7516/2019/10/076
[arXiv: 1909.03034 [gr-qc]].

\bibitem{Burda:2015isa} P.~Burda, R.~Gregory and I.~Moss,
``Gravity and the stability of the Higgs vacuum,''
Phys. Rev. Lett. \textbf{115} (2015) 071303,
doi: 10.1103/PhysRevLett.115.071303
[arXiv: 1501.04937 [hep-th]].

\bibitem{Burda:2015yfa} P.~Burda, R.~Gregory and I.~Moss,
``Vacuum metastability with black holes,''
JHEP \textbf{08} (2015) 114,
doi: 10.1007/JHEP08(2015)114
[arXiv: 1503.07331 [hep-th]].

\bibitem{Tetradis:2016vqb} N.~Tetradis,
``Black holes and Higgs stability,''
JCAP \textbf{09} (2016) 036,
doi: 10.1088/1475-7516/2016/09/036
[arXiv: 1606.04018 [hep-ph]].

\bibitem{Gorbunov:2017fhq} D.~Gorbunov, D.~Levkov and A.~Panin,
``Fatal youth of the Universe: black hole threat for the electroweak vacuum during preheating,''
JCAP \textbf{10} (2017) 016,
doi: 10.1088/1475-7516/2017/10/016
[arXiv: 1704.05399 [astro-ph.CO]].

\bibitem{Canko:2017ebb} D.~Canko, I.~Gialamas, G.~Jelic-Cizmek, A.~Riotto and N.~Tetradis,
``On the catalysis of the electroweak vacuum decay by black holes at high temperature,''
Eur. Phys. J. C \textbf{78} (2018) 328,
doi: 10.1140/epjc/s10052-018-5808-y
[arXiv: 1706.01364 [hep-th]].

\bibitem{Mukaida:2017bgd} K.~Mukaida and M.~Yamada,
``False vacuum decay catalyzed by black holes,''
Phys. Rev. D \textbf{96} (2017) 103514,
doi: 10.1103/PhysRevD.96.103514
[arXiv: 1706.04523 [hep-th]].

\bibitem{Kohri:2017ybt} K.~Kohri and H.~Matsui,
``Electroweak vacuum collapse induced by vacuum fluctuations of the Higgs field around evaporating black holes,''
Phys. Rev. D \textbf{98} (2018) 123509,
doi: 10.1103/PhysRevD.98.123509
[arXiv: 1708.02138 [hep-ph]].

\bibitem{Dai:2019eei} D.~C.~Dai, R.~Gregory and D.~Stojkovic,
``Connecting the Higgs potential and primordial black holes,''
Phys. Rev. D \textbf{101} (2020) 125012,
doi: 10.1103/PhysRevD.101.125012
[arXiv: 1909.00773 [hep-ph]].


\bibitem{tHooft:1984kcu} G.~'t Hooft,
``On the quantum structure of a black hole,''
Nucl. Phys. B \textbf{256} (1985) 727,
doi: 10.1016/0550-3213(85)90418-3.

\bibitem{Ghosh:2009xg} K.~Ghosh,
``Near-horizon geometry and the entropy of a minimally coupled scalar field in the Schwarzschild black hole,''
J. Phys. Soc. Jap. \textbf{85} (2016) 014101,
doi: 10.7566/JPSJ.85.014101
[arXiv: 0902.1601 [gr-qc]].

\bibitem{brick1} S.~Mukohyama and W.~Israel,
``Black holes, brick walls and the Boulware state,''
Phys. Rev. D \textbf{58} (1998) 104005,
doi: 10.1103/PhysRevD.58.104005
[arXiv: gr-qc/9806012 [gr-qc]].

\bibitem{brick2} S.~Banerjee, S.~Das, M.~Dorband and A.~Kundu,
``Brickwall, normal modes, and emerging thermality,''
Phys. Rev. D \textbf{109} (2024) 126020,
doi: 10.1103/PhysRevD.109.126020
[arXiv: 2401.01417 [hep-th]].


\bibitem{imbh1} J.~E.~Greene, J.~Strader and L.~C.~Ho,
``Intermediate-mass black holes,''
Ann. Rev. Astron. Astrophys. \textbf{58} (2020) 257,
doi: 10.1146/annurev-astro-032620-021835
[arXiv: 1911.09678 [astro-ph.GA]].


\bibitem{smbh1} L.~Ferrarese and H.~Ford,
``Supermassive black holes in galactic nuclei: Past, present and future research,''
Space Sci. Rev. \textbf{116} (2005) 523,
doi: 10.1007/s11214-005-3947-6
[arXiv: astro-ph/0411247 [astro-ph]].

\bibitem{smbh2} K.~Gultekin, \textit{et al.}
``The M-sigma and M-L relations in galactic bulges and determinations of their intrinsic scatter,''
Astrophys. J. \textbf{698} (2009) 198,
doi: 10.1088/0004-637X/698/1/198
[arXiv: 0903.4897 [astro-ph.GA]].

\bibitem{smbh3} J.~Kormendy and L.~C.~Ho,
``Coevolution (or not) of supermassive black holes and host galaxies,''
Ann. Rev. Astron. Astrophys. \textbf{51} (2013) 511,
doi: 10.1146/annurev-astro-082708-101811
[arXiv: 1304.7762 [astro-ph.CO]].


\bibitem{Blandford:2018iot} R.~Blandford, D.~Meier and A.~Readhead,
``Relativistic jets from active galactic nuclei,''
Ann. Rev. Astron. Astrophys. \textbf{57} (2019) 467,
doi: 10.1146/annurev-astro-081817-051948
[arXiv: 1812.06025 [astro-ph.HE]].

\bibitem{Jets} S.~Komissarov and O.~Porth,
``Numerical simulations of jets,''
New. Astr. Rev. \textbf{92} (2021) 101610,
doi: 10.1016/j.newar.2021.101610.


\bibitem{jwst1} I.~Labbe, \textit{et al.}
``A population of red candidate massive galaxies \textasciitilde{}600 Myr after the big bang,''
Nature \textbf{616} (2023) 266,
doi: 10.1038/s41586-023-05786-2
[arXiv: 2207.12446 [astro-ph.GA]].

\bibitem{Harikane:2022rqt} Y.~Harikane, {\it et al.}
``A Comprehensive study of galaxies at z \ensuremath{\sim} 9\textendash{}16 found in the early JWST data: Ultraviolet luminosity functions and cosmic star formation history at the pre-reionization epoch,''
Astrophys. J. Suppl. \textbf{265} (2023) 5,
doi: 10.3847/1538-4365/acaaa9
[arXiv: 2208.01612 [astro-ph.GA]].

\bibitem{Maiolino:2023bpi} R.~Maiolino, \textit{et al.}
``JADES. The diverse population of infant black holes at 4\ensuremath{<}z\ensuremath{<}11: Merging, tiny, poor, but mighty,''
[arXiv: 2308.01230 [astro-ph.GA]].

\bibitem{Greene:2024phl} J.~E.~Greene, \textit{et al.}
``UNCOVER spectroscopy confirms the surprising ubiquity of active galactic nuclei in red sources at z \ensuremath{>} 5,''
Astrophys. J. \textbf{941} (2022) 106,
doi: 10.3847/1538-4357/ad1e5f.


\bibitem{jwstSMBH1} B.~Liu and V.~Bromm,
``Accelerating early massive galaxy formation with primordial black holes,''
Astrophys. J. Lett. \textbf{937} (2022) L30,
doi: 10.3847/2041-8213/ac927f
[arXiv: 2208.13178 [astro-ph.CO]].


\bibitem{smbhBeforeJWST1} X.~Fan, E.~Banados and R.~A.~Simcoe,
``Quasars and the intergalactic medium at cosmic dawn,''
Annu. Rev. Astron. Astrophys.\textbf{61} (2023) 373,
doi: 10.1146/annurev-astro-052920-102455
[arXiv: 2212.06907 [astro-ph.GA]].

\bibitem{smbhJWST1} R.~L.~Larson \textit{et al.} [CEERS Team],
``A CEERS discovery of an accreting supermassive black hole 570 Myr after the big bang: Identifying a progenitor of massive z \ensuremath{>} 6 quasars,''
Astrophys. J. Lett. \textbf{953} (2023) L29,
doi: 10.3847/2041-8213/ace619
[arXiv: 2303.08918 [astro-ph.GA]].

\bibitem{smbhJWST2} H.~\"Ubler, \textit{et al.}
``GA-NIFS: JWST discovers an offset AGN 740 million years after the big bang,''
Mon. Not. Roy. Astron. Soc. \textbf{531} (2024) 355,
doi: 10.1093/mnras/stae943
[arXiv: 2312.03589 [astro-ph.GA]].

\bibitem{smbhJWST3} M.~Onoue, \textit{et al.}
``A Candidate for the least-massive black hole in the first 1.1 billion years of the universe,''
Astrophys. J. Lett. \textbf{942} (2023) L17,
doi: 10.3847/2041-8213/aca9d3
[arXiv: 2209.07325 [astro-ph.GA]].

\bibitem{smbhJWST4} R.~Maiolino, \textit{et al.}
``A small and vigorous black hole in the early Universe,''
Nature \textbf{627} (2024) 59,
[erratum: Nature \textbf{630} (2024) E2]
doi: 10.1038/s41586-024-07494-x
[arXiv: 2305.12492 [astro-ph.GA]].

\bibitem{Xiao} M.~Xiao, \textit{et al.}
``Accelerated formation of ultra-massive galaxies in the first billion years,''
Nature \textbf{635} (2024) 311,
doi: 10.1038/s41586-024-08094-5.

\bibitem{smbhJWST5} A.~Bogdan, \textit{et al.}
``Evidence for heavy-seed origin of early supermassive black holes from a z\,\ensuremath{\approx}\,10 X-ray quasar,''
Nature Astron. \textbf{8} (2024) 126,
doi: 10.1038/s41550-023-02111-9
[arXiv: 2305.15458 [astro-ph.GA]].

\bibitem{smbhJWST6} L.~J.~Furtak, \textit{et al.}
``A high black hole to host mass ratio in a lensed AGN in the early Universe,''
Nature \textbf{628} (2024) 57,
doi: 10.1038/s41586-024-07184-8
[arXiv: 2308.05735 [astro-ph.GA]].

\bibitem{smbhJWST7} O.~E.~Kovacs, \textit{et al.}
``A Candidate supermassive black hole in a gravitationally lensed galaxy at z \ensuremath{\approx} 10,''
Astrophys. J. Lett. \textbf{965} (2024) L21,
doi: 10.3847/2041-8213/ad391f
[arXiv: 2403.14745 [astro-ph.GA]].


\bibitem{mg1} C.~D.~Bailyn, R.~K.~Jain, P.~Coppi and J.~A.~Orosz,
``The Mass distribution of stellar black holes,''
Astrophys. J. \textbf{499} (1998) 367,
doi: 10.1086/305614
[arXiv: astro-ph/9708032 [astro-ph]].

\bibitem{mg2} F.~Ozel, D.~Psaltis, R.~Narayan and J.~E.~McClintock,
``The Black hole mass distribution in the galaxy,''
Astrophys. J. \textbf{725} (2010) 1918,
doi: 10.1088/0004-637X/725/2/1918
[arXiv: 1006.2834 [astro-ph.GA]].

\bibitem{mg3} W.~M.~Farr, N.~Sravan, A.~Cantrell, L.~Kreidberg, C.~D.~Bailyn, I.~Mandel and V.~Kalogera,
``The Mass distribution of stellar-mass black holes,''
Astrophys. J. \textbf{741} (2011) 103,
doi: 10.1088/0004-637X/741/2/103
[arXiv: 1011.1459 [astro-ph.GA]].

\bibitem{mg4} K.~Belczynski, G.~Wiktorowicz, C.~Fryer, D.~Holz and V.~Kalogera,
``Missing Black Holes unveil the supernova explosion mechanism,''
Astrophys. J. \textbf{757} (2012) 91,
doi: 10.1088/0004-637X/757/1/91
[arXiv: 1110.1635 [astro-ph.GA]].

\bibitem{Adler:1969gk} S.~L.~Adler,
``Axial vector vertex in spinor electrodynamics,''
Phys. Rev. \textbf{177} (1969) 2426,
doi: 10.1103/PhysRev.177.2426.

\bibitem{Bell:1969ts} J.~S.~Bell and R.~Jackiw,
``A PCAC puzzle: $\pi^0 \to \gamma \gamma$ in the $\sigma$ model,''
Nuovo Cim. A \textbf{60} (1969) 47,
doi: 10.1007/BF02823296.

\bibitem{tHooft:1976rip} G.~'t Hooft,
``Symmetry breaking through Bell-Jackiw anomalies,''
Phys. Rev. Lett. \textbf{37} (1976) 8,
doi: 10.1103/PhysRevLett.37.8.

\bibitem{Kuzmin:1985mm} V.~A.~Kuzmin, V.~A.~Rubakov and M.~E.~Shaposhnikov,
``On the anomalous electroweak baryon number nonconservation in the early universe,''
Phys. Lett. B \textbf{155} (1985) 36,
doi: 10.1016/0370-2693(85)91028-7.

\bibitem{Fukugita:1986hr} M.~Fukugita and T.~Yanagida,
``Baryogenesis without grand unification,''
Phys. Lett. B \textbf{174} (1986) 45,
doi: 10.1016/0370-2693(86)91126-3.

\bibitem{Shaposhnikov:1987tw} M.~E.~Shaposhnikov,
``Baryon asymmetry of the universe in standard electroweak theory,''
Nucl. Phys. B \textbf{287} (1987) 757,
doi: 10.1016/0550-3213(87)90127-1.

\bibitem{Lan-Lif} L.~D.~Landau and E.~M.~Lifschitz,
                  {\it The Classical Theory of Fields}
                  (Pergamon Press, Oxford 1975).

\bibitem{Manton:1983nd} N.~S.~Manton,
``Topology in the Weinberg-Salam theory,''
Phys. Rev. D \textbf{28} (1983) 2019,
doi: 10.1103/PhysRevD.28.2019.

\bibitem{Klinkhamer:1984di} F.~R.~Klinkhamer and N.~S.~Manton,
``A Saddle point solution in the Weinberg-Salam theory,''
Phys. Rev. D \textbf{30} (1984) 2212,
doi: 10.1103/PhysRevD.30.2212.

\bibitem{DeLuca:2021oer} V.~De Luca, G.~Franciolini, A.~Kehagias and A.~Riotto,
``Standard model baryon number violation seeded by black holes,''
Phys. Lett. B \textbf{819} (2021) 136454,
doi: 10.1016/j.physletb.2021.136454
[arXiv: 2102.07408 [astro-ph.CO]].

\bibitem{Gregory:2013hja} R.~Gregory, I.~G.~Moss and B.~Withers,
``Black holes as bubble nucleation sites,''
JHEP \textbf{03} (2014) 081,
doi: 10.1007/JHEP03(2014)081
[arXiv: 1401.0017 [hep-th]].

\bibitem{Coleman:1980aw} S.~R.~Coleman and F.~De Luccia,
``Gravitational effects on and of vacuum decay,''
Phys.\ Rev.\ D {\bf 21} (1980) 3305,
doi: 10.1103/PhysRevD.21.3305.

\bibitem{Mellor:1989wc} F.~Mellor and I.~Moss,
``Black holes and gravitational instantons,''
Class. Quant. Grav. \textbf{6} (1989) 1379,
doi: 10.1088/0264-9381/6/10/008.

\bibitem{Dowker:1993bt} F.~Dowker, J.~P.~Gauntlett, D.~A.~Kastor and J.~H.~Traschen,
``Pair creation of dilaton black holes,''
Phys. Rev. D \textbf{49} (1994) 2909,
doi: 10.1103/PhysRevD.49.2909
[arXiv: hep-th/9309075 [hep-th]].

\bibitem{Hawking:1998bn} S.~W.~Hawking and N.~Turok,
``Open inflation without false vacua,''
Phys. Lett. B \textbf{425} (1998) 25,
doi: 10.1016/S0370-2693(98)00234-2
[arXiv: hep-th/9802030 [hep-th]].

\bibitem{Coleman:1977} S.~R.~Coleman,
``The Fate of the false cacuum. 1. Semiclassical theory,''
Phys. Rev. D \textbf{15} (1977) 2929,
[erratum: Phys. Rev. D \textbf{16} (1977) 1248]
doi: 10.1103/PhysRevD.16.1248.


\bibitem{BH-1} S.~Chandrasekhar,
              {\it The Mathematical Theory of Black Holes}
              (Springer, Clarendon, New York 1983).

\bibitem{BH-2} S.~Carroll,
              {\it Spacetime and Geometry: An Introduction to General Relativity}
              (Addison-Wesley, San Francisco 2004).

\bibitem{BH-3} E.~Poisson,
              {\it A Relativist's Toolkit: The Mathematics of Black-Hole Mechanics}
              (Cambridge University Press, Cambridge 2004).

\bibitem{Wald} R.~M.~Wald,
               {\it General Relativity}
               (University of Chicago Press, Chicago 1984).

\bibitem{gravMTW} C.~W.~Misner, K.~S.~Thorne, J.~A.~Wheeler,
                 {\it Gravitation}
                 (Freeman, San Francisco 1973).

\bibitem{meG4} M.~Gogberashvili,
``Einstein's hole argument and Schwarzschild singularities,''
Annals Phys. \textbf{452} (2023) 169274,
doi: 10.1016/j.aop.2023.169274
[arXiv: 2303.10348 [gr-qc]].

\bibitem{Colombeau-1} J.~F.~Colombeau,
                      {\it New Generalized Functions and Multiplication of Distributions}
                      ((Eslsevier, North Holland, Amsterdam 1984).

\bibitem{Colombeau-2} J.~F.~Colombeau,
                      {\it Elementary Introduction to New Generalized Functions}
                      (Eslsevier, North Holland, Amsterdam 1985).

\bibitem{Gel-Sch} I.~M.~Gelfand and G.~E.~Schilov,
                  {\it Generalized Functions. Vol. I: Properties and Operations}
                  (Academic Press, New York - London 1964).

\bibitem{Pantoja:1997zt} N.~R.~Pantoja and H.~Rago,
``Energy-momentum tensor valued distributions for the Schwarzschild and Reissner-Nordstrom geometries,''
[arXiv: gr-qc/9710072 [gr-qc]].

\bibitem{Heinzle:2001bk} J.~M.~Heinzle and R.~Steinbauer,
``Remarks on the distributional Schwarzschild geometry,''
J. Math. Phys. \textbf{43} (2002) 1493,
doi: 10.1063/1.1448684
[arXiv: gr-qc/0112047 [gr-qc]].

\bibitem{Foukzon} J.~Foukzon, A.~Potapov and E.~Menkova,
``Was Polchinski wrong? Colombeau distributional Rindler space-time with distributional Levi-Civit\'a connection induced vacuum dominance. Unruh effect revisited,''
JHEPGC \textbf{4} (2018) 361,
doi: 10.4236/jhepgc.2018.42023.

\bibitem{Foukzon1} J.~Foukzon, E.~R.~Men'kova, A.~A.~Potapov and S.~A.~Podosenov,
``Was Polchinski wrong? Colombeau distributional Rindler space-time with distributional Levi-Civit\`a connection induced vacuum dominance. Unruh effect revisited,''
J. Phys. Conf. Ser. \textbf{1141} (2018) 012100,
doi: 10.1088/1742-6596/1141/1/012100.

\bibitem{Modif-1} S.~B.~Giddings,
``Black hole information, unitarity, and nonlocality,''
Phys. Rev. D \textbf{74} (2006) 106005,
doi: 10.1103/PhysRevD.74.106005
[arXiv: hep-th/0605196 [hep-th]].

\bibitem{Modif-3} J.~Maldacena and L.~Susskind,
``Cool horizons for entangled black holes,''
Fortsch. Phys. \textbf{61} (2013) 781,
doi: 10.1002/prop.201300020
[arXiv: 1306.0533 [hep-th]].

\bibitem{Kerr:2023rpn} R.~P.~Kerr,
``Do black holes have singularities?,''
[arXiv: 2312.00841 [gr-qc]].

\bibitem{Cardoso:2019rvt} V.~Cardoso and P.~Pani,
``Testing the nature of dark compact objects: a status report,''
Living Rev. Rel. \textbf{22} (2019) 4,
doi: 10.1007/s41114-019-0020-4
[arXiv: 1904.05363 [gr-qc]].

\bibitem{Ansoldi:2008jw} S.~Ansoldi,
``Spherical black holes with regular center: A Review of existing models including a recent realization with Gaussian sources,''
[arXiv: 0802.0330 [gr-qc]].

\bibitem{Malafarina:2017csn} D.~Malafarina,
``Classical collapse to black holes and quantum bounces: A review,''
Universe \textbf{3} (2017) 48,
doi: 10.3390/universe3020048
[arXiv: 1703.04138 [gr-qc]].

\bibitem{meG1} M.~Gogberashvili and L.~Pantskhava,
``Black hole information problem and wave bursts,''
Int. J. Theor. Phys. \textbf{57} (2018) 1763,
doi: 10.1007/s10773-018-3702-x
[arXiv: 1608.04595 [physics.gen-ph]].

\bibitem{meG2} M.~Gogberashvili,
``Can quantum particles cross a horizon?,''
Int. J. Theor. Phys. \textbf{58} (2019) 3711,
doi: 10.1007/s10773-019-04242-0
[arXiv: 1712.02637 [gr-qc]].

\bibitem{meG3} R.~Beradze, M.~Gogberashvili and L.~Pantskhava,
``Reflective black holes,''
Mod. Phys. Lett. A \textbf{36} (2021) 2150200,
doi: 10.1142/S021773232150200X.

\bibitem{Howard:1984qp} K.~W.~Howard and P.~Candelas,
``Quantum stress tensor in Schwarzschild space-time,''
Phys. Rev. Lett. \textbf{53} (1984) 403,
doi: 10.1103/PhysRevLett.53.403.

\bibitem{Mathur:2005zp} S.~D.~Mathur,
``The Fuzzball proposal for black holes: An Elementary review,''
Fortsch. Phys. \textbf{53} (2005) 793,
doi: 10.1002/prop.200410203
[arXiv: hep-th/0502050 [hep-th]].

\bibitem{Modif-2} A.~Almheiri, D.~Marolf, J.~Polchinski and J.~Sully,
``Black holes: Complementarity or firewalls?,''
JHEP \textbf{02} (2013) 062,
doi: 10.1007/JHEP02(2013)062
[arXiv: 1207.3123 [hep-th]].

\bibitem{Einstein} A.~Einstein,
``On a stationary system with spherical symmetry consisting of many gravitating masses,''
Ann. Math. \textbf{40} (1939) 922.

\bibitem{Schwarz} K.~Schwarzschild,
``On the gravitational field of a mass point according to Einstein’s theory,''
Sitzungsber. Preuss. Akad. Wiss., Phys. Math. Kl. (1916) 189.


\bibitem{Chitishvili:2021jrx} M.~Chitishvili, M.~Gogberashvili, R.~Konoplich and A.~S.~Sakharov,
``Higgs field-induced triboluminescence in binary black hole mergers,''
Universe \textbf{9} (2023) 301,
doi: 10.3390/universe9070301
[arXiv: 2111.07178 [astro-ph.HE]].

\bibitem{Col-Wei} S.~R.~Coleman and E.~J.~Weinberg,
``Radiative corrections as the origin of spontaneous symmetry breaking,''
Phys. Rev. D {\bf 7} (1973) 1888,
doi: 10.1103/PhysRevD.7.1888.

\bibitem{Peskin_Book} M.~E.~Peskin and D.~V.~Schroeder,
                     {\it An Introduction to Quantum Field Theory}
                     (Perseus Books, Reading 1995).

\bibitem{Schwartz:2013pla} M.~D.~Schwartz,
                          {\it Quantum Field Theory and the Standard Model}
                          (Cambridge University Press, Cambridge 2014).

\bibitem{Tye:2016pxi} S.~H.~H.~Tye and S.~S.~C.~Wong,
``The Chern-Simons number as a dynamical variable,''
Ann. Math. Sci. Appl. \textbf{01} (2016) 123,
doi: 10.4310/amsa.2016.v1.n1.a3
[arXiv: 1601.00418 [hep-th]].

\bibitem{Zh-Ho} M.-Y.~Zhuang and L.~C.~Ho,
``Evolutionary paths of active galactic nuclei and their host galaxies,''
Nature Astronomy \textbf{7} (2023) 1376,
doi: 10.1038/s41550-023-02051-4
[arXiv: 2308.08603 [astro-ph.GA]].

\bibitem{Ricarte:2023owr} A.~Ricarte, R.~Narayan and B.~Curd,
``Recipes for jet feedback and spin evolution of black holes with strongly magnetized super-Eddington accretion disks,''
Astrophys. J. Lett. \textbf{954} (2023) L22,
doi: 10.3847/2041-8213/aceda5
[arXiv: 2307.04621 [astro-ph.HE]].

\bibitem{EHTC} K.~Akiyama, {\it et al.} [Event Horizon Telescope Collaboration],
``First Sagittarius A* Event Horizon Telescope Results. VIII. Physical interpretation of the polarized ring,''
Astrophys. J. Lett. \textbf{964} (2024) L26,
doi: 10.3847/2041-8213/ad2df1.

\bibitem{Inayoshi:2019fun} K.~Inayoshi, E.~Visbal and Z.~Haiman,
``The assembly of the first massive black holes,''
Ann. Rev. Astron. Astrophys. \textbf{58} (2020) 27,
doi: 10.1146/annurev-astro-120419-014455
[arXiv: 1911.05791 [astro-ph.GA]].

\bibitem{Volonteri:2021sfo} M.~Volonteri, M.~Habouzit and M.~Colpi,
``The origins of massive black holes,''
Nature Rev. Phys. \textbf{3} (2021) 732,
doi: 10.1038/s42254-021-00364-9
[arXiv: 2110.10175 [astro-ph.GA]].


\bibitem{tree01} M.~Volonteri, G.~Lodato and P.~Natarajan,
``The evolution of massive black hole seeds,''
Mon. Not. Roy. Astron. Soc. \textbf{383} (2008) 1079,
doi: 10.1111/j.1365-2966.2007.12589.x
[arXiv: 0709.0529 [astro-ph]].

\bibitem{tree1} F.~Marulli, S.~Bonoli, E.~Branchini, L.~Moscardini and V.~Springel,
``Modeling the cosmological co-evolution of supermassive black holes and galaxies. 1. BH scaling relations and the AGN luminosity function,''
Mon. Not. Roy. Astron. Soc. \textbf{385} (2008) 1846,
doi: 10.1111/j.1365-2966.2008.12988.x
[arXiv: 0711.2053 [astro-ph]].

\bibitem{tree2} P.~Dayal, E.~M.~Rossi, B.~Shiralilou, O.~Piana, T.~R.~Choudhury and M.~Volonteri,
``The hierarchical assembly of galaxies and black holes in the first billion years: predictions for the era of gravitational wave astronomy,''
Mon. Not. Roy. Astron. Soc. \textbf{486} (2019) 2336,
doi: 10.1093/mnras/stz897
[arXiv: 1810.11033 [astro-ph.GA]].

\bibitem{tree3} O.~Piana, P.~Dayal, M.~Volonteri and T.~R.~Choudhury,
``The mass assembly of high-redshift black holes,''
Mon. Not. Roy. Astron. Soc. \textbf{500} (2020) 2146,
doi: 10.1093/mnras/staa3363
[arXiv: 2009.13505 [astro-ph.GA]].

\bibitem{tree4} A.~Trinca, R.~Schneider, R.~Valiante, L.~Graziani, L.~Zappacosta and F.~Shankar,
``The low-end of the black hole mass function at cosmic dawn,''
Mon. Not. Roy. Astron. Soc. \textbf{511} (2022) 616,
doi: 10.1093/mnras/stac062
[arXiv: 2201.02630 [astro-ph.GA]].

\bibitem{tree5} A.~Toubiana, {\it et al.}
``Reconciling PTA and JWST and preparing for LISA with POMPOCO: a Parametrisation Of the Massive black hole POpulation for Comparison to Observations,''
[arXiv: 2410.17916 [astro-ph.GA]].

\bibitem{hydro1} D.~J.~Croton, {\it et al.}
``The Many lives of AGN: Cooling flows, black holes and the luminosities and colours of galaxies,''
Mon. Not. Roy. Astron. Soc. \textbf{365} (2006) 11,
[erratum: Mon. Not. Roy. Astron. Soc. \textbf{367} (2006) 864]
doi: 10.1111/j.1365-2966.2006.09994.x
[arXiv: astro-ph/0602065 [astro-ph]].

\bibitem{hydro2} J.~Schaye, \textit{et al.}
``The EAGLE project: Simulating the evolution and assembly of galaxies and their environments,''
Mon. Not. Roy. Astron. Soc. \textbf{446} (2015) 521,
doi: 10.1093/mnras/stu2058
[arXiv: 1407.7040 [astro-ph.GA]].

\bibitem{hydro3} D.~Sijacki, {\it et al.}
``The Illustris simulation: the evolving population of black holes across cosmic time,''
Mon. Not. Roy. Astron. Soc. \textbf{452} (2015) 575,
doi: 10.1093/mnras/stv1340
[arXiv: 1408.6842 [astro-ph.GA]].

\bibitem{eps} J.~Ellis, M.~Fairbairn, J.~Urrutia and V.~Vaskonen,
``What is the origin of the JWST SMBHs?,''
[arXiv: 2410.24224 [astro-ph.CO]].

\bibitem{Pacucci:2023oci} F.~Pacucci, B.~Nguyen, S.~Carniani, R.~Maiolino and X.~Fan,
``JWST CEERS and JADES active galaxies at z = 4\textendash{}7 violate the local M$_\bullet$ - M$_{{\ensuremath{\star}}}$ relation at \ensuremath{>}3\ensuremath{\sigma}: Implications for low-mass black holes and seeding models,''
Astrophys. J. Lett. \textbf{957} (2023) L3,
doi: 10.3847/2041-8213/ad0158
[arXiv: 2308.12331 [astro-ph.GA]].

\bibitem{Juodzbalis} F.~Pacucci, A.~Loeb and I.~Juod\v{z}balis,
``The Host galaxy of a dormant, overmassive black hole at z\,=\,6.7 may be restarting star formation,''
Res. Notes AAS \textbf{8} (2024) 105,
doi: 10.3847/2515-5172/ad3fb8
[arXiv: 2404.11643 [astro-ph.GA]].



\bibitem{bProd1} T.~Asaka, D.~Grigoriev, V.~Kuzmin and M.~Shaposhnikov,
``Late reheating, hadronic jets and baryogenesis,''
Phys. Rev. Lett. \textbf{92} (2004) 101303,
doi: 10.1103/PhysRevLett.92.101303
[arXiv: hep-ph/0310100 [hep-ph]].


\bibitem{symRate1} P.~B.~Arnold and L.~D.~McLerran,
``Sphalerons, small fluctuations and baryon number violation in electroweak theory,''
Phys. Rev. D \textbf{36} (1987) 581,
doi: 10.1103/PhysRevD.36.581

\bibitem{symRate2} P.~B.~Arnold, D.~Son and L.~G.~Yaffe,
``The Hot baryon violation rate is O (alpha-w**5 T**4),''
Phys. Rev. D \textbf{55} (1997) 6264,
doi: 10.1103/PhysRevD.55.6264
[arXiv: hep-ph/9609481 [hep-ph]]

\bibitem{symRate3} G.~D.~Moore,
``Sphaleron rate in the symmetric electroweak phase,''
Phys. Rev. D \textbf{62} (2000) 085011,
doi: 10.1103/PhysRevD.62.085011
[arXiv: hep-ph/0001216 [hep-ph]].

\bibitem{Tew1} M.~D'Onofrio, K.~Rummukainen and A.~Tranberg,
``Sphaleron rate in the minimal standard model,''
Phys. Rev. Lett. \textbf{113} (2014) 141602,
doi: 10.1103/PhysRevLett.113.141602
[arXiv: 1404.3565 [hep-ph]]


\bibitem{CPinSM1} M.~B.~Gavela, P.~Hernandez, J.~Orloff and O.~Pene,
``Standard model CP violation and baryon asymmetry,''
Mod. Phys. Lett. A \textbf{9} (1994) 795,
doi: 10.1142/S0217732394000629
[arXiv: hep-ph/9312215 [hep-ph]].

\end{thebibliography}
\end{document}